\documentclass[10pt]{iopart}
\usepackage{iopams}
\usepackage{dcolumn}
\usepackage{graphicx}

\usepackage[slovak,english]{babel}
\usepackage[T1]{fontenc}
\usepackage[utf8]{inputenc}
\usepackage{color}

\begin{document}
\title{The magnetic field induced ferromagnetism in EuPd$_2$Sn$_4$ novel compound}
\author{I \v{C}url\'ik$^1$, M Zapotokov\'a$^1$, F Gastaldo$^2$, M Reiffers$^{1,3}$, J G Sereni$^4$, M Giovannini$^2$}
\address{$^1$Faculty of Humanities and Natural Sciences, University of Pre\v{s}ov, 17. novembra 1, Pre\v{s}ov, Slovakia}
\address{$^2$Department of Chemistry, University of Genova, Via Dodecaneso 31, Genova, Italy}
\address{$^3$Institute of Experimental Physics, Slovak Academy of Science, Watsonova 47, Ko\v{s}ice, Slovakia}
\address{$^4$Department of Physics, CAB-CNEA, CONICET, IB-UNCuyo, 8400 S. C. de Bariloche, Argentina}
\ead{ivan.curlik@unipo.sk}

\begin{abstract}
{We report on crystal structure, magnetic and thermal physical properties of EuPd$_2$Sn$_4$ stannide. From the magnetic susceptibility measurements a divalent state of Eu rare earth element was determined together with an antiferromagnetic (AFM) order at 11 K. By applying magnetic field, the magnetic behavior variation reveals a complex magnetic structure. Above a critical field $B_c$ a ferromagnetic (FM)-like character starts to prevail. AFM and FM regimes are separated by a metamagnetic region. Heat capacity measurements confirm this behavior.}
\end{abstract}
  
\maketitle

\section{Introduction}
Eu, as well as Sm, Tm, Yb and Ce intermetallic compounds constitute an interesting class of materials, owing to different configuration of their $4f$-electrons which can lead to the phenomenon of valence instability where valence fluctuates between integer values \cite{Lawrence_1981}.

In the last years we have carried out a systematic investigation on $R$-Pd-$X$ ( $R$ = Eu, Yb, Ce; $X$ = In, Sn) ternary compounds among which a certain number of them exhibited anomalous physical behaviour \cite{Muramatsu_2011, Giovannini_2010, Sereni_2011, Martinelli_2019}.

Concerning the Eu-Pd-Sn system, novel compounds have been discovered, such as Eu$_3$Pd$_2$Sn$_2$ and EuPdSn$_2$, showing a stable Eu$^{2+}$ magnetic state and complex magnetic structures ~\cite{Solokha_2011, Curlik_2017, Curlik_2018}. This last feature, together with a magnetic anisotropy, seems to characterize many compounds belonging to Eu-$T$-$X$ systems ($T$ = transition metals; $X$ = $p$-block elements). In fact, magnetic properties measured on a single crystal of EuRh$_2$Si$_2$ exhibit a marked magnetic anisotropy with the easy plane perpendicular to the $c$ axis ~\cite{Seiro_2013}, whereas a single crystal study on EuRhGe$_3$ revealed significant anisotropy and incommensurate antiferromagnetic structures with amplitude-modulated character \cite{Bednarchuk_2015_JALCOM}. These elaborated magnetic scenarios require to be verified by microscopic techniques, like neutron diffraction. A careful neutron scattering study was indeed performed on EuPdSn, confirming the complex magnetic structure occurring in these compounds. In fact, EuPdSn exhibits a non collinear sine-wave modulated structure below $T = 15.5$~K which evolves into a planar helimagnetic structure at lower temperatures \cite{Lemoine_2012}.

These scenarios of complex and strongly anisotropic magnetism are not expected for Eu$^{2+}$ because of its spin $S=7/2$ and orbital $L = 0$ numbers precluding the presence of crystal electric field effects.

Following the same line of study of Eu-based compounds, in this work we report on the physical properties of EuPd$_2$Sn$_4$, a new ternary member of the Eu-Pd-Sn system.

\section{Experimental details}
EuPd$_2$Sn$_4$ polycrystalline sample has been prepared by weighing the stoichiometric amount of elements with the following nominal purity: Eu – 99.99 mass~\% (pieces, Smart Elements GmbH, Vienna, Austria), Pd – 99.95 mass~\% (foil, Chimet, Arezzo, Italy), Sn – 99.999 mass~\% (bar, Smart Elements GmbH, Vienna, Austria). Due to Eu element being prone to high oxidation reaction on air, the EuPd$_2$Sn$_4$ sample has been weighed inside a glove box. In order to avoid the loss of europium during the melting because of its high vapour pressure, all the weighed elements (in total weight of 1.2~g) were enclosed in a small tantalum crucible sealed by arc welding in an inert atmosphere inside the glove box.
 
The sample was subsequently melted in an induction furnace under a stream of pure argon. To ensure homogeneity, the crucible was continuously shaken during the melting. Sample was then annealed in a resistance furnace for two weeks at the temperature of $600~^\circ$C and finally quenched in cold water.

The sample was characterized by scanning electron microscopy (EVO 40, Carl Zeiss, Cambridge, England) equipped with quantitative electron probe microanalysis system based on energy dispersive X-ray spectroscopy (EPMA - EDXS). For the quantitative and qualitative analysis an acceleration voltage of 20~keV for 100~s was applied, and a cobalt standard was used for calibration. The crystal structure was determined by powder X-ray diffraction (XRD) using the vertical diffractometer X$\prime$Pert MPD (Philips, Almelo, The Netherlands) equipped with a graphite monochromator installed in the diffraction beam (Bragg Brentano, Cu K$\alpha$ radiation). The theoretical powder pattern was calculated with the help of the Powder-Cell software ~\cite{Kraus_1996} and for Rietveld refinements the FULLPROFF program was used ~\cite{Rodriguez_1993}.

The thermodynamic and magnetic physical properties measurements were performed using a Physical Property Measurement System Dynacool (Quantum Design) in the temperature range of 2 – 400~K under applied magnetic field up to 9~T. Specific heat was determined by means of the relaxation 2-$\tau$ method.

\section{Experimental results}

\subsection{Crystal structure analysis}
From SEM/EPMA analyses the sample prepared on nominal composition EuPd$_2$Sn$_4$ resulted to be almost single phase with some traces of Eu$_3$Pd$_4$Sn$_{13}$ (crystallizing in the Yb$_3$Rh$_4$Sn$_{13}$ structure type) and an unknown phase of atomic composition 5~\% Eu, 40~\% Pd, 55~\%Sn.

\begin{figure}
  \centering
  \includegraphics[width=8.2cm]{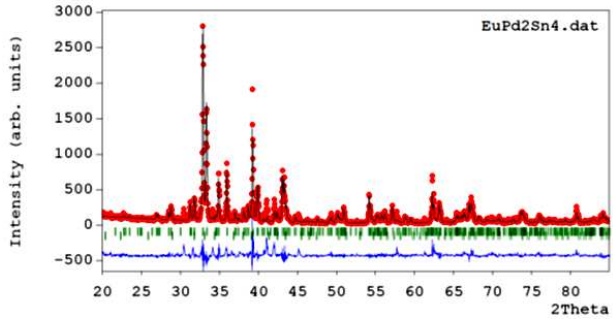}
  \caption{X–ray diffraction powder pattern of EuPd$_2$Sn$_4$. The experimental data are shown with circle symbols, while the solid line through the experimental points represents Rietveld refinement. The indexed peaks associated to the crystal structure of EuPd$_2$Sn$_4$ are displayed with the upper ticks (lower ticks indicate traces of Eu$_3$Pd$_4$Sn$_{13}$). The lower curve represents the difference curve.}
  \label{Fig.1}
\end{figure}

The XRD pattern of the EuPd$_2$Sn$_4$ phase (shown in Fig.~\ref{Fig.1}) was successfully indexed by analogy with the corresponding already known orthorhombic phase NdAu$_2$In$_4$ ~\cite{Salvador_2007}, which crystallizes in NdRh$_2$Sn$_4$-structure type ($oP28$) ~\cite{Meot-Meyer_1985}. Its crystal structure was established as orthorhombic, space group \textit{Pnma} (N\textsuperscript{\underline{o}}. 62).

The atomic positions of NdAu$_2$In$_4$ were taken as starting values and structure was successfully refined in the XRD dataset by using FULLPROF software \cite{Rodriguez_1993}. The refined lattice parameters for EuPd$_2$Sn$_4$ are: $a=1.8659(7)$~nm, $b=0.4600(6)$~nm, $c=0.7284(5)$~nm. The details of the structural atomic coordinates of EuPd$_2$Sn$_4$ are reported in Tab.~\ref{Tab.I}. A final reliability factor of $R_p=22.6~\%$ and $\chi^2=3.71$ were obtained in the Rietveld refinement (see Fig.~\ref{Fig.1}).

\begin{table}
\caption{\label{Tab.I}Atomic fractional coordinates of the novel compound EuPd$_2$Sn$_4$.}
\begin{indented}
\item[]\begin{tabular}{@{}ccccc}
\br
  \textbf{Atom}&\textbf{Wyckoff position}&\textbf{\textit{x/a}}&\textbf{\textit{y/b}}&\textbf{\textit{z/c}}\\
\mr
  Sn1&4$c$&0.0316(3)&1/4&0.366(1)\\
  Sn2&4$c$&0.1817(4)&1/4&0.816(1)\\
  Sn3&4$c$&0.1966(5)&1/4&0.235(1)\\
  Sn4&4$c$&0.4327(4)&1/4&0.486(1)\\
  Pd1&4$c$&0.0346(5)&1/4&0.741(1)\\
  Pd2&4$c$&0.2913(5)&1/4&0.538(1)\\
  Eu&4$c$&0.3601(5)&1/4&0.009(1)\\
\br
\end{tabular}
\end{indented}
\end{table}

\subsection{Magnetic properties}
\begin{figure}
  \centering
  \includegraphics[width=8.2cm]{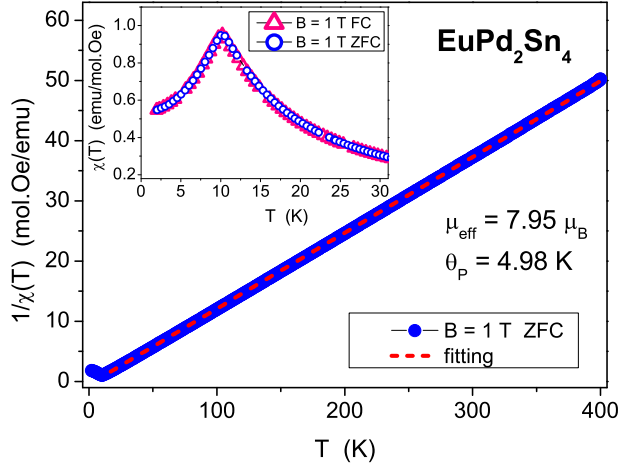}
  \caption{Temperature dependence $1/\chi(T)$ for EuPd$_2$Sn$_4$ compound measured in the temperature range 2 – 400~K in applied magnetic field of 1 T. In the inset the $\chi(T)$ dependence in the temperature range 2 – 30~K with the maximum at 11~K is shown.}
  \label{Fig.2}
\end{figure}

Fig.~\ref{Fig.2} shows the temperature dependence of the inverse magnetic susceptibility $1/\chi(T)$ of EuPd$_2$Sn$_4$ in the temperature range of $2-400$~K measured in the applied magnetic field of 1~T. The compound exhibits Curie-Weiss (CW) behaviour above 11~K. To evaluate the basic magnetic characteristics we fitted $1/\chi(T)$ dependence in this temperature range following the equation
\begin{equation}
\label{Eq.1}
1/\chi(T)={\left[\chi_0+\frac{C}{T-\Theta_P}\right]}^{-1}
\end{equation}
where $\chi_0$, $C$ and $\Theta_P$ represent the Pauli susceptibility, Curie constant and paramagnetic Curie temperature, respectively. The obtained value of the effective magnetic moment $\mu_{eff}=7.95~\mu_B$ is close to the theoretical Eu$^{2+}$ free ion value (7.94~$\mu_B$) indicating that Eu ions are in the magnetic state. The small but positive value of the paramagnetic Curie temperature $\theta_P=4.98$~K denotes some ferromagnetic (FM) exchange interactions between magnetic moments. The Pauli susceptibility $\chi_0$ obtained from the fit is negligible ($\sim 10^{-10}$~m$^3$mol$^{-1}$). For comparison, we fitted also the $1/\chi(T)$ dependence in applied magnetic field 9~T. The resulting values ($\theta_P=3.34$~K, $\mu_{eff}=7.98~\mu_B$) reveal negligible changes in comparison to those obtained from measurements in 1~T.

The inset of Fig.~\ref{Fig.2} presents the low temperature details of the $\chi(T)$ dependence of EuPd$_2$Sn$_4$ with a peak at $T_N=11$~K indicating a phase transition from paramagnetic to magnetically ordered state. Notably, such a sharp cusp is characteristic for antiferromagnetic (AFM) ordering. As one can see there is no obvious difference between zero field cooling (ZFC) and field cooling (FC) dependencies (in analogy with the results observed in EuPdSn$_2$ \cite{Curlik_2018}) suggesting an AFM order. However, taking into account the positive $\theta_P$ value, a more complex magnetic structure can not be excluded. Complex magnetic structures were already described in EuPdSn, where neutron diffraction experiment revealed incommensurate magnetic structures (sinusoidally modulated vs. planar helimagnetic) in different temperature ranges well below the $T_N$ transition \cite{Lemoine_2012}.

\begin{figure}
  \centering
  \includegraphics[width=8.2cm]{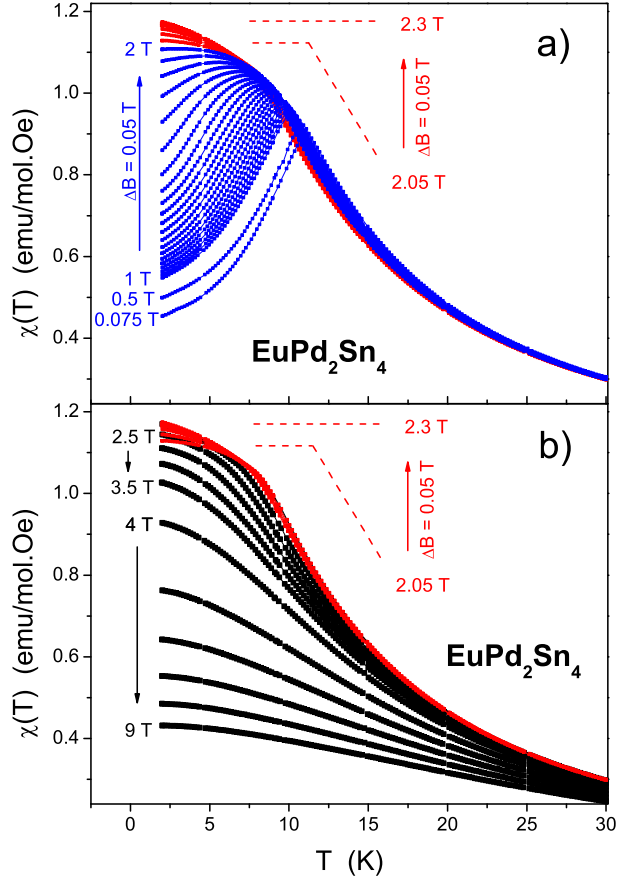}
  \caption{Set of temperature dependent $\chi(T)$ susceptibilities for EuPd$_2$Sn$_4$ compound measured in the temperature range $2-30$~K in magnetic fields within the range $0.075-2.3$~T (panel $a$) and $2.05-9$~T (panel $b$). According to the characteristic AFM/FM-like shape, the $\chi(T)$ dependencies are depicted on the corresponding $a/b$ panel of the figure.}
  \label{Fig.3}
\end{figure}

In order to proceed with a detailed investigation of the ground-state magnetic properties of EuPd$_2$Sn$_4$, a series of $\chi(T)$ measurements were performed focusing to the low temperature range between $2-30$~K in different applied magnetic fields up to 9~T. The $\chi(T)$ dependencies were measured in FC regime where magnetic field was gradually increased by steps of $0.05$~T for every consecutive measurement. Resulted dependencies are presented on $a$ and $b$ panels of Fig.~\ref{Fig.3}. For applied fields up to around 1~T, $\chi(T)$ shows a cusp characteristic for AFM ordering (the blue curves in Fig. \ref{Fig.3}a). With increasing $B$ up to 2~T, the transition temperature $T_N$ shifts to lower values and the peak broadens. For $B>2$~T the $\chi(T)$ dependencies change their character, the AFM maximum disappears and $\chi(T)$ continuously increases with lowering the temperature (black curves on Fig.~\ref{Fig.3}b). The displayed set of $\chi(T)$ measurements shows the trend of magnetic saturation effect with increasing $B$. It seems that FM-like correlations become prevalent in the system. This could explain also the positive $\theta_P$ calculated from the fitting. Within the $B$ range of $\approx2-2.3$~T (red curves in both $a, b$ panels of Fig.~\ref{Fig.3}) the absolute values of $\chi(T)$ measurements are progressively higher and this is typical for a progressive FM field induced alignment from an AFM state. A similar switching of prevailing type of magnetic interactions between AFM and FM was found in EuPdSn$_2$ measurements \cite{Curlik_2018}, whereas antiferromagnetic EuRh$_2$Si$_2$ was found to be on the verge of ferromagnetism above barely 0.1~T ~\cite{Hossain_2001, Seiro_2011}.

\begin{figure}
  \centering
  \includegraphics[width=8.2cm]{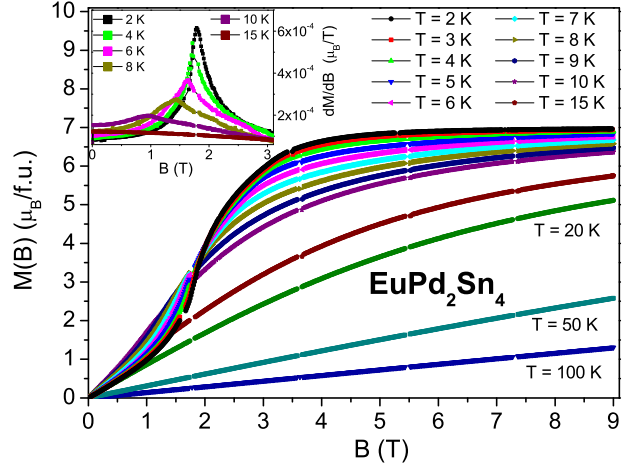}
  \caption{Isothermal $M(B)$ magnetization of EuPd$_2$Sn$_4$ for the magnetic field $0-9$~T measured at selected temperatures within the range $2-100$~K. Inset of the figure shows the field dependencies of $dM$/$dB$ derivative in the $0-3$~T range for selected isotherms.}
  \label{Fig.4}
\end{figure}

This scenario of field-induced ferromagnetism in EuPd$_2$Sn$_4$ is corroborated by the $M(B)$ magnetization dependencies at temperatures from 2 to 100~K (Fig.~\ref{Fig.4}). In the $M(B)$ isotherms measured above the ordering temperature $T_N=11$~K, the curves show the typical field dependence described by a Brillouin function expected for a paramagnetic compound, whereas below $T_N$ a metamagnetic-like behavior in the $M(B)$ takes place for the values of critical magnetic field $B_c$ between 1 and 2~T. The $B_c$ was evaluated as the maximum of $dM$/$dB$ derivative (see the inset of Fig.~\ref{Fig.4}), which represents the magnetic field for which the AFM regime is destroyed and the metamagnetic processes predominate. The inset of Fig.~\ref{Fig.4} shows the maxima corresponding to $B_c$ for different $M(B)$ isotherms where, with increasing temperature, $B_c$ assumes smaller values (depicted later in Fig.~\ref{Fig.5} with orange pentagons). Far above the metamagnetic transition at $B=9$~T and $T = 2$ K the magnetization saturates at 6.95 $\mu_B$/Eu which is very close to the free Eu$^{2+}$ saturation moment $gJ = 7~\mu_B$/Eu.

\subsection{Specific heat}
\begin{figure}
  \centering
  \includegraphics[width=8.2cm]{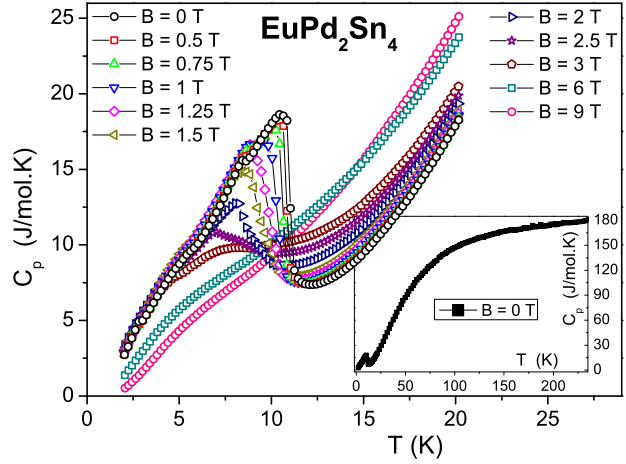}
  \caption{Low temperature dependence of specific heat $C_p(T)$ of EuPd$_2$Sn$_4$ in different magnetic field from the range $0-9$~T. The inset presents the $C_p(T)$ measurements in zero field up to high temperatures.}
  \label{Fig.6}
\end{figure}

The specific heat $C_p(T)$ of EuPd$_2$Sn$_4$ in the temperature range $2-230$~K for zero magnetic field is reported in the inset of Fig.~\ref{Fig.6}. In the low temperature region a $\lambda$-anomaly at $\approx11$~K is observed, indicating the transition into AFM state. The main panel of Fig.~\ref{Fig.6} presents the detail of this anomaly in accordance to different applied magnetic fields up to 9~T. For magnetic field $B < 2$~T the $C_p(T_N)$ jump starts being reduced both in the temperature and the height. This behavior is a typical feature of antiferromagnetically ordered systems in agreement with the $\chi(T)$ results (see Fig.~\ref{Fig.3}). In comparison with the magnetic measurements, the $T_N$ values extracted from $C_p(T)$ dependencies equally match with the Neél temperatures obtained from the $\chi(T)$ and tends to the same AFM trend below $B_c$. The $C_p(T)$ dependencies on Fig.~\ref{Fig.6} also confirm the progressive FM-like character for $B>2.5$~T where the maximum shifts to higher $T$ and broadens with the strength of $B$ (characteristic for FM polarization).
At lower temperature (around 5~K) a shoulder-like anomaly is observed. Such a broad shoulder, seen around 1/4 of $T_N$, is typical for $4f^7$ systems (Eu$^{2+}$ and Gd$^{3+}$) and is related to the large degeneracy of the $J=7/2$ local moment within the mean-field theory for a ($2J+1$)-fold degenerate multiplet ~\cite{Curlik_2018, Seiro_2011, Bednarchuk_2015_APPa}.

\subsection{Magnetic phase diagram}
\begin{figure}
  \centering
  \includegraphics[width=8.2cm]{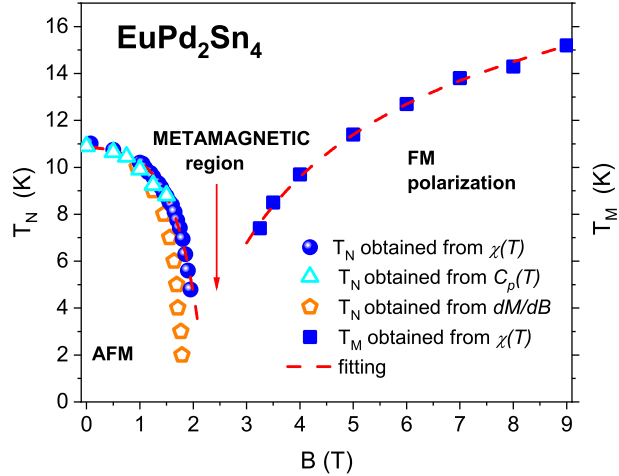}
  \caption{The magnetic phase diagram shows the variation of transition temperature $T_N$ and characteristic temperature $T_M$ as a function of external magnetic field $B$. The graphics displays the variation of dominant magnetic interactions in EuPd$_2$Sn$_4$ as a function of applied magnetic field.}
  \label{Fig.5}
\end{figure}

Fig.~\ref{Fig.5} presents the analysis of the transition temperature as a function of applied magnetic field from the series of $\chi(T)$ curves (Fig.~\ref{Fig.3}), $M(B)$ magnetization (Fig.~\ref{Fig.4}) and the $C_p(T)$ specific heat measurements (Fig.~\ref{Fig.6}). For values of magnetic field ranging from zero up to $\approx2$~T the $T_N$ values, taken from maxima of $\chi(T)$ curves, decrease in agreement with typical AFM ordering (blue circles). The same trend of $T_N$ variation with magnetic field results from the analysis of $C_p(T)$ dependencies measured at $B\leq1.5$~T (cyan triangles). For magnetic fields above $2.5$~T, the magnetic evolution of the system can be described tracing the temperature of the maximum $d\chi/dT$ slope: $T_{M(B)} = (d\chi/dT)_{max}$ (blue squares in Fig.~\ref{Fig.5}). This field induced ferromagnetic behaviour is consistent with the isothermal $M(B)$ magnetization measurements (inset of Fig.~\ref{Fig.4}), where the critical magnetic field $B_c$ was evaluated for different temperatures. In Fig.~\ref{Fig.5} we present this $T_N$ vs. $B_c$ dependence (orange pentagons). The graphics is divided in two areas depending on the dominating type of interaction (AFM vs FM). On the magnetic field axis the AFM/FM-like regimes are separated by the region of metamagnetic behaviour left bounded by the critical magnetic field $B_c$. Such an explanation congruent with the $\chi(T)$ susceptibility (Fig.~\ref{Fig.3}) and $M(B)$ magnetization (Fig.~\ref{Fig.4}) measurements is supported also by the results for the analogous system – EuPtIn where two metamagnetic transitions between AFM and FM spin alignment were observed ~\cite{Mullmann_1998}.

\section{Conclusions}
The EuPd$_2$Sn$_4$ compound has been synthesised by induction melting, and magnetic and thermal physical properties have been studied. In this stannide Eu was found in divalent state, which makes EuPd$_2$Sn$_4$ a magnetic compound. The series of $\chi(T)$ magnetic susceptibilities measured in low magnetic field reveal an antiferromagnetic ordering at 11~K, but increasing the strength of $B$ a more complex magnetic nature was observed. A typical AFM ground state was observed for values lower than $B < 2$~T and a further increase of magnetic field induces FM-like behaviour. These two regimes are separated by a metamagnetic region revealed by $M(B)$ measurements. The complexity and variation of magnetic mechanisms in EuPd$_2$Sn$_4$ are supported by specific heat experiment where $C_p(T)$ dependencies in $B > 2.5$~T are characterized by field induced FM behaviour, possibly due to a non-collinear ordering of magnetic moments in AFM state. As a consequence, the increasing of magnetic field $B$ would cause the re-arrangement of magnetic moments towards a FM-like state. However, a change in the magnetic structure should be confirmed by neutron powder diffraction measurements.

\ack{Acknowledgements}
This work was supported by the projects VEGA 1/0956/17, VEGA 1/0611/18, VEGA 1/0705/20 and APVV-16-0079, University Science Park TECHNICOM for Innovation Applications Supported by Knowledge Technology - II. phase (ITMS: 313011D232), Research \& Development Operational Programme funded by the ERDF.

\section*{References}
\bibliography{reference}

\providecommand{\newblock}{}
\begin{thebibliography}{10}
\expandafter\ifx\csname url\endcsname\relax
  \def\url#1{{\tt #1}}\fi
\expandafter\ifx\csname urlprefix\endcsname\relax\def\urlprefix{URL }\fi
\providecommand{\eprint}[2][]{\url{#2}}

\bibitem{Lawrence_1981}
Lawrence J~M, Riseborough P~S and Parks R~D 1981 {\em Reports on Progress in
  Physics\/} {\bf 44} 1--84

\bibitem{Muramatsu_2011}
Muramatsu T, Kanemasa T, Kagayama T, Shimizu K, Aoki Y, Sato H, Giovannini M,
  Bonville P, Zlatic V, Aviani I, Khasanov R, Rusu C, Amato A, Mydeen K,
  Nicklas M, Michor H and Bauer E 2011 {\em Physical Review B\/} {\bf 83} ISSN
  1098-0121

\bibitem{Giovannini_2010}
Giovannini M, Pasero R and Saccone A 2010 {\em Intermetallics\/} {\bf 18} 429
  -- 433

\bibitem{Sereni_2011}
Sereni J~G, Giovannini M, Berisso M~G and Saccone A 2011 {\em Physical Review
  B\/} {\bf 83}

\bibitem{Martinelli_2019}
Martinelli A, Sanna S, Lamura G, Ritter C, Joseph B, Bauer E and Giovannini M
  2019 {\em Journal of Physics: Condensed Matter\/} {\bf 31}

\bibitem{Solokha_2011}
Solokha P, \v{C}url\'ik I, Giovannini M, Lee-Hone N, Reiffers M, Ryan D and
  Saccone A 2011 {\em Journal of Solid State Chemistry\/} {\bf 184} 2498--2505

\bibitem{Curlik_2017}
\v{C}url\'ik I, Gastaldo F, Giovannini M, Strydom A and Reiffers M 2017 {\em
  Acta Physica Polonica A\/} {\bf 131} 1003--1005

\bibitem{Curlik_2018}
{\v{C}}url{\'{\i}}k I, Giovannini M, Gastaldo F, Strydom A~M, Reiffers M and
  Sereni J~G 2018 {\em Journal of Physics: Condensed Matter\/} {\bf 30} 495802

\bibitem{Seiro_2013}
Seiro S and Geibel C 2013 {\em Journal of Physics: Condensed Matter\/} {\bf 26}
  046002

\bibitem{Bednarchuk_2015_JALCOM}
Bednarchuk O and Kaczorowski D 2015 {\em Journal of Alloys and Compounds\/}
  {\bf 646} 291 -- 297 ISSN 0925-8388

\bibitem{Lemoine_2012}
Lemoine P, Cadogan J~M, Ryan D~H and Giovannini M 2012 {\em Journal of Physics:
  Condensed Matter\/} {\bf 24} 236004

\bibitem{Kraus_1996}
Kraus W and Nolze G 1996 {\em Journal of Applied Crystallography\/} {\bf 29}
  301--303

\bibitem{Rodriguez_1993}
Rodr\'iguez-Carvajal J 1993 {\em Physica B: Condensed Matter\/} {\bf 192} 55 --
  69 ISSN 0921-4526

\bibitem{Salvador_2007}
Salvador J~R, Hoang K, Mahanti S~D and Kanatzidis M~G 2007 {\em Inorganic
  Chemistry\/} {\bf 46} 6933--6941

\bibitem{Meot-Meyer_1985}
Méot-Meyer M, Venturini G, Malaman B and Roques B 1985 {\em Materials Research
  Bulletin\/} {\bf 20} 913 -- 919 ISSN 0025-5408

\bibitem{Hossain_2001}
Hossain Z, Trovarelli O, Geibel C and Steglich F 2001 {\em Journal of Alloys
  and Compounds\/} {\bf 323-324} 396 -- 399 ISSN 0925-8388 proceedings of the
  4th International Conference on f-Elements

\bibitem{Seiro_2011}
Seiro S and Geibel C 2011 {\em Journal of Physics: Condensed Matter\/} {\bf 23}
  375601

\bibitem{Bednarchuk_2015_APPa}
Bednarchuk O and Kaczorowski D 2015 {\em Acta Physica Polonica A\/} {\bf 127}
  418 -- 420

\bibitem{Mullmann_1998}
Müllmann R, Mosel B~D, Eckert H, Kotzyba G and Pöttgen R 1998 {\em Journal of
  Solid State Chemistry\/} {\bf 137} 174 -- 180 ISSN 0022-4596

\end{thebibliography}
\bibliographystyle{iopart-num}

\end{document}